\documentstyle[12pt]{article}

\title{Three-Dimensional Billiards with Time 
Machine\footnote{Published in Intern. J. Mod. Phys. D5, 179-192 (1996)}}
\author{\large Michael B.Mensky$^{*\#\dag}$ and
Igor D.Novikov$^{\#\S\P\copyright}$\\[10pt]
$^*$P.N.Lebedev Physical Institute, 117924 Moscow, Russia\\
Email: mensky@sci.lebedev.ru\\[3pt]
$^{\#}$Theoretical Astrophysics Center, Blegdamsvej 17\\
DK-2100 Copenhagen {\O}, Denmark\\[3pt]
$^{\dag}$Institut f\"ur Theoretische Physik,
Technische Universit\"at Berlin\\
Hardenbergstr. 36,
D-10623 Berlin, Germany\\[3pt]
$^{\S}$NORDITA, Blegdamsvej 17\\
DK21-00 Copenhagen {\O}, Denmark\\[3pt]
$^{\P}$Astro Space Center, P.N.Lebedev Physical Institute\\
Profsoyuznaya 84/32, Moscow 117810, Russia\\[3pt]
$^{\copyright}$University Observatory, {\O}ster Voldgade 3\\
DK-1350 Copenhagen K, Denmark
}

\date{}

\newcommand{\be}{\begin{equation}}
\newcommand{\ee}{\end{equation}}
\newcommand{\ba}{\begin{eqnarray}}
\newcommand{\ea}{\end{eqnarray}}
\newcommand{\ban}{\begin{eqnarray*}}
\newcommand{\ean}{\end{eqnarray*}}

\newcommand{\r}{{\mathbf r}}
\renewcommand{\u}{{\mathbf u}}
\renewcommand{\d}{{\mathbf d}}
\renewcommand{\L}{{\mathbf L}}
\renewcommand{\l}{{\mathbf l}}
\newcommand{\rone}{{\mathbf r}_1}
\newcommand{\rtwo}{{\mathbf r}_2}
\newcommand{\vone}{{\mathbf v}_1}
\newcommand{\vtwo}{{\mathbf v}_2}

\newcommand{\cone}{{\mathbf c}_1}
\newcommand{\ctwo}{{\mathbf c}_2}
\renewcommand{\c}{{\mathbf c}}
\newcommand{\ri}{{\mathbf r}_i}
\newcommand{\rf}{{\mathbf r}_f}
\newcommand{\rA}{{\mathbf r}_A}
\newcommand{\rB}{{\mathbf r}_B}
\newcommand{\rzer}{{\mathbf r}_0}

\newcommand{\lb}{\lambda_b}
\newcommand{\lw}{\lambda_w}
\newcommand{\lone}{\lambda_1}
\newcommand{\ltwo}{\lambda_2}
\newcommand{\D}{{\mathbf D}}

\begin{document}
\maketitle \newpage
{Self-collision of a non-relativistic classical point-like body,
or particle, in the spacetime containing closed time-like curves
(time-machine spacetime) is considered. A point-like body
(particle) is an idealization of a small ideal elastic billiard
ball. The known model of a time machine is used containing a
wormhole leading to the past. If the body enters one of the
mouths of the wormhole, it emerges from another mouth in an
earlier time so that both the particle and its ``incarnation''
coexist during some time and may collide. Such self-collisions
are considered in the case when the size of the body is much less
than the radius of the mouth, and the latter is much less than
the distance between the mouths. Three-dimensional configurations
of trajectories with a self-collision are presented.  Their
dynamics is investigated in detail. Configurations corresponding
to multiple wormhole traversals are discussed. It is shown that,
for each world line describing self-collision of a particle,
dynamically equivalent configurations exist in which the particle
collides not with itself but with an identical particle having a
closed trajectory (Jinnee of Time Machine).}

\vspace{0.5cm}
PACS number(s): 04.20Cv, 04.20.Ex, 98.80.Hw.
\vspace{0.5cm}

\section{Introduction}\label{intro}

Spacetimes containing closed time-like curves ({\em time-machine
spacetimes}) were considered in the literature in the last decade
(see \cite{Th91}-\cite{Nov89} and references therein). We shall
investigate dynamics of a point-like body (small billiard ball)
in a time-machine spacetime, taking into account
a self-collision of a particle. We use a wide-known model of the
time-machine spacetime containing a wormhole. In such a model a
body entering one of the mouths of the wormhole, goes out of
another mouth at an earlier time.

In a series of recent papers (see for example
\cite{Fr90}-\cite{Nov93}) the action of the standard laws of
physics is investigated in different time-machine spacetimes.
Specifically, it was argued in \cite{MinAct} that the principle
of self-consistency for motion in time-machine spacetimes
(introduced earlier in \cite{Nov-self1}-\cite{Nov-self2}) may be
derived from the principle of minimal action. Therefore, if the
initial and final positions of all bodies are given, then the
motion in the time-machine spacetime is completely determined by
the principle of minimal action. Of course, it could happen that
there are more than one local minimum of the action or even
infinite number of them. Then different ways of motion are
possible corresponding to different local minima.

In the present paper we shall apply this approach to investigate
the motion of a non-relativistic classical point-like body (which
will be called a particle) in the time-machine spacetime
containing a wormhole. The point-like body may be thought of as
an idealization of a small ideal elastic billiard ball. We
restrict ourselves by the case when $\rho\ll r\ll l$ where
$\rho$, $r$ and $l$ are correspondingly the radius of the ball,
the radius of the mouths of the time machine and the distance
between the mouths.

Complanar (and close to complanar) motions in a time-machine
spacetime with a self-collision were analyzed in \cite{Ech91}
for billiard balls of a finite size and in \cite{MinAct}, from a
different point of view, for a particle, i.e. infinitesimal
billiard ball. In the present paper we shall investigate
three-dimensional motion of a particle with a
self-collision.\footnote{Possibility of three-dimensional motions
was mentioned also in \cite{MinAct} though not analyzed there in
detail.} Concrete examples will be given of the motion with a
self-collision and repeated passages through the time machine.

\section{The method and a simple solution}\label{method}

To be definite, we shall consider the time-machine spacetime
containing the wormhole with two mouths, $A$ and $B$. If
something (for example a particle) enters the mouth $B$ at time
$t_B$, it immediately (counting in the proper time) goes out of
the mouth $A$, but its time in the external spacetime turns out
to be $t_A=t_B-\tau$. The parameter $\tau$ characterizes
``power'' of the time machine.

During some time before entering the mouth $B$ the particle coexists
with its ``incarnation'' escaped from the mouth $A$, and an
interaction between them is possible during this period. We shall
consider only the case of collision, i.e. short-distance
interaction. Three-dimensional motion of the particle including
such a self-collision and wormhole traversal will be considered
in Sect.~\ref{one-travers}. Later, in Sect.~\ref{multiple-pass},
we shall discuss the case of multiple traversals.

Following the paper \cite{MinAct}, we shall derive configuration
of the process starting with the principle of minimal action,
that guarantees self-consistency of the motion.

\begin{figure}
%%Begin InstantTeX Picture
\let\picnaturalsize=N
\def\picsize{7cm}
\def\picfilename{les1.eps}
%If you do not have the picture file add:
%\let\nopictures=Y
%to the beginning of the file.
\ifx\nopictures Y\else{\ifx\epsfloaded Y\else\input epsf \fi
\let\epsfloaded=Y
\centerline{\ifx\picnaturalsize N\epsfxsize \picsize\fi \epsfbox{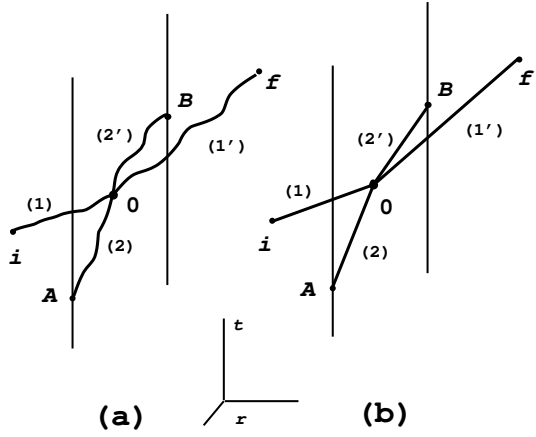}}}\fi
%% End InstantTeX Picture
\caption{\rm The trajectory with a self-collision and one passage
through the wormhole: (a)~the action principle is not accounted;
(b)~the action principle is accounted between the key events.}
\label{traject}\end{figure}

The trajectory of the particle begins (see Fig.~\ref{traject}a)
at the space-time point (event) $i=(\ri, t_i)$ and ends at the
event $f=(\rf, t_f)$. The collision occurs at the event
$O=(\rzer, t_0)$. Let us denote the entities characterizing the
motion during the first period, between the events $i$ and $O$,
by letters with the index $1$. For example, position during this
period will be denoted by $\rone(t)$. The entities corresponding
to the last period, from the event $O$ to the event $f$, will be
denoted also by the letters having the index $1$ and prime.
For example, position during this period will be denoted
$\rone'(t)$.\footnote{These notations are not generally accepted,
but they proved to be convenient for our aims.}

After collision at the event $(\rzer, t_0)$ the particle enters
the wormhole mouth $B$ at the event $(\rB, t_B)$. On escaping
from the mouth $A$ at the event ($\rA, t_A$) the particle travels
to the event of collision, ($\rzer, t_0$). The entities between
the events ($\rA, t_A$) and ($\rzer, t_0$) will be denoted by the
index $2$, and those between ($\rzer, t_0$) and ($\rB, t_B$) by
the index $2$ and prime (correspondingly $\rtwo(t)$ and
$\rtwo'(t)$ for the positions).

The action of the particle may be calculated as a sum of the
actions corresponding to the mentioned parts of the
trajectory:\footnote{A short-range interaction (collision) does
not contribute the action. The proof of this for a specific form
of the interaction potential may be found in \cite{MinAct}, but
the same is valid for much more wide class of potentials.}
\be\label{action}
S=\frac{m}{2}\left( \int\dot\rone^2 dt + \int\dot\rone'^2 dt +
 \int\dot\rtwo^2 dt + \int\dot\rtwo'^2 dt\right).
\ee
Small variation of the part $1$ of the trajectory (between the
events ($\ri, t_i$) and ($\rzer, t_0$)) does not change the
action (\ref{action}) only if the motion is linear in this part.
Quite analogously, the requirement that the action must be
stationary separately for the part $2$, or the part $1'$, or the
part $2'$ of the trajectory, leads to the conclusion that the
motion is linear for each of these parts.

Therefore, the trajectory has the form shown in
Fig.~\ref{traject}b, and the action is given by the following
rather simple formula:
\be\label{action-v}
S= \frac{m}{2}\left[ \vone^2(t_0-t_i) + \vone'^2(t_f-t_0)
+\vtwo^2(t_0-t_A) +\vtwo'^2(t_B-t_0)\right]
\ee
where the following notations are introduced for the constant
velocities in the parts of the trajectory between the key events:
\be\label{vel-events}
\vone=\frac{\rzer-\ri}{t_0-t_i}, \quad
\vone'=\frac{\rf-\rzer}{t_f-t_0}, \quad
\vtwo=\frac{\rzer-\rA}{t_0-t_A}, \quad
\vtwo'=\frac{\rB-\rzer}{t_B-t_0}.
\ee

The principle of minimal action is not yet accounted completely.
The boundary conditions $\ri$, $t_i$, $\rf$, $t_f$ as well as the
locations of the mouths $\rA$ and $\rB$ and the time shift
$t_B-t_A=\tau$ are fixed, but the event of collision, ($\rzer,
t_0$), as well as the sum $t_A+t_B$, may be chosen arbitrarily.
The principle of minimal action requires that a small variation
of these parameters do not alter the action. The problem of
minimum of the functional is reduced thus to the problem of
minimum of the function of finite number of parameters.

\begin{figure}
%%Begin InstantTeX Picture
\let\picnaturalsize=N
\def\picsize{7cm}
\def\picfilename{les2.eps}
%If you do not have the picture file add:
%\let\nopictures=Y
%to the beginning of the file.
\ifx\nopictures Y\else{\ifx\epsfloaded Y\else\input epsf \fi
\let\epsfloaded=Y
\centerline{\ifx\picnaturalsize N\epsfxsize \picsize\fi \epsfbox{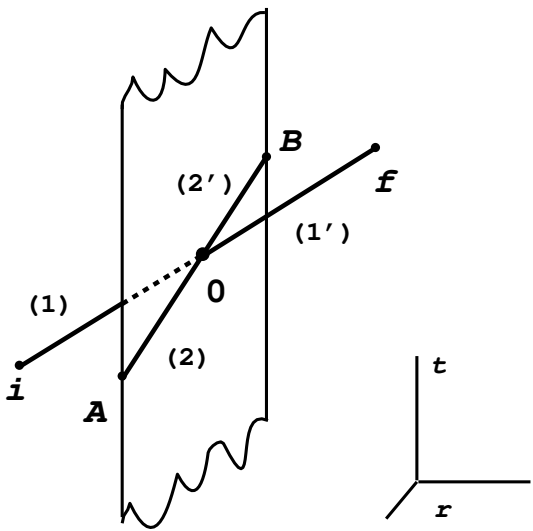}}}\fi
%%End InstantTeX Picture
\caption{\rm Simple trajectory minimizing the action: a collision
with exchange of velocities.}
\label{trivial}\end{figure}

There is one evident solution of this problem. To find it, one
should consider the parts $1$ and $1'$ of the trajectory jointly,
as a single motion, and the parts $2$ and $2'$ as one more
motion. The action has a minimum when each of these motions is
linear, i.e.
$$
\vone=\vone', \quad \vtwo=\vtwo'
$$
(see Fig.~\ref{trivial}). Indeed, any small change of the
parameters $\rzer$, $t_0$ and $t_A+t_B$ converts the straight
line connecting the events $i$, $f$ and/or the straight line
connecting $(t_A,\rA)$, $(t_B,\rB)$ into broken lines, increasing
the action (\ref{action-v}).

This configuration of the trajectory corresponds to the case when
the velocities of the particle and its incarnation are exchanged
in the collision.  Of course, this solution exists only if the
direct line (in the spacetime) connecting the events ($\ri, t_i$)
and ($\rf, t_f$) intersects (see Fig.~\ref{trivial}) the part of
the plane between the world lines of the mouths, $\rA$ (arbitrary
time) and $\rB$ (arbitrary time). The velocity $\vtwo=\vtwo'$ in
this case is directed from the mouth $A$ to the mouth $B$ and is
equal to the distance between the mouths divided by $\tau$.

This simple configuration is complanar in the sense that all key
points $\ri$, $\rf$, $\rA$, $\rB$ and $\rzer$ lie in the same
plane. Of course, this configuration is well known
\cite{Ech91,MinAct} and is mentioned here only as an illustration
of the method.

\section{General solution for one traversal}\label{one-travers}

Variation of the action (\ref{action-v}) (with the notations
(\ref{vel-events})) in the parameters $\rzer$, $t_0$ and
$t_A+t_B$ gives the equations\footnote{They have been
derived by
this method in
\cite{MinAct}}
\be\label{vel-eq}
\vone + \vtwo = \vone' + \vtwo', \quad
\vone^2+\vtwo^2 = \vone'^2+\vtwo'^2, \quad
\vtwo^2 = \vtwo'^2
\ee
The first and the second of these equations express correspondingly 
conservation of the momentum and energy in the collision. The third 
equation gives the rule that the particle exits from the time machine 
(wormhole) with the same velocity that it has had at the entrance.

It is easily seen that the simple motion considered in the
preceding section (see Fig.~\ref{trivial}) is a solution of the
equations (\ref{vel-eq}). The general solution has also very
simple form
\cite{MinAct}
\be\label{vel-sol}
\vone = \cone - \c, \quad \vone' = \cone + \c, \quad
\vtwo = \ctwo + \c, \quad \vtwo' = \ctwo - \c, \quad
\ee
where the vectors $\cone$, $\ctwo$ and $\c$ satisfy the condition
\be\label{orthog}
\cone\, \c =0, \quad \ctwo\, \c =0
\ee
and are otherwise arbitrary.

If we know all the velocities, we can find all parameters of the
motion (the spacetime points determining the trajectories of
Fig.~\ref{traject}b). It is more difficult to solve the opposite
problem and find $\cone$, $\ctwo$, $\c$ given the characteristics
of the time machine $\rA$, $\rB$, $\tau$ and boundary conditions
($\ri$, $t_i$), ($\rf$, $t_f$).  This will be our task in the
present section.

This problem may be in principle solved if one considers
Eq.~(\ref{vel-eq}) (with the notations (\ref{vel-events})) as
equations not for velocities, but for the unknown space-time
parameters $\rzer$, $t_0$ and $t_A+t_B$. Indeed, substituting the
expressions from (\ref{vel-events}) for the velocities in
Eq.~(\ref{vel-eq}), we obtain one vector and two scalar equations
for one vector and two scalar unknowns.

We shall slightly modify this scheme using the parametrization
(\ref{vel-sol}) for the velocities. Substituting the expressions
(\ref{vel-sol}) for the velocities in (\ref{vel-events}) and
performing elementary operations with the resulting equalities,
we have
\ba
T\,\cone+T_b\,\c=\L, \quad
\tau \,\ctwo - T_w\,\c=\l, \quad
T_b\,\cone-T_w\,\ctwo+(T+\tau)\,\c=\d \label{main-eq}\\
4\rzer=\ri+\rf+\rA+\rB-T_b\,\cone-T_w\,\ctwo-(T-\tau)\,\c.
\label{rzer-eq}\ea
where the following notations are introduced (``$b$'' stands for
``boundary'' and ``$w$'' for ``wormhole''):
\ban
\L=\rf-\ri, \quad
\l=\rB-\rA, \quad
\d=(\ri+\rf)-(\rA+\rB), \\
T=t_f-t_i, \quad
T_b=(t_i+t_f-2t_0), \quad
T_w=(t_A+t_B-2t_0).
\ean

Now we shall accept the following strategy. The linear equations
(\ref{main-eq}) for the vectors $\cone$, $\ctwo$, $\c$ may be
easily solved giving the expressions of these vectors through the
unknowns $T_b$ and $T_w$. Then the orthogonality conditions
(\ref{orthog}) give equations for $T_b$ and $T_w$. Finally
$\rzer$ may be found from Eq.~(\ref{rzer-eq}).

Before realizing this program, let us consider a simple special
case
\be\label{special}
T_b=0, \quad T_w=0.
\ee
In this case we have (from (\ref{main-eq}))
$$
\cone=\frac{\L}{T}, \quad
\ctwo=\frac{\l}{\tau}, \quad
\c=\frac{\d}{T+\tau}.
$$
The orthogonality conditions (\ref{orthog}) should be imposed,
and this means that this special case is realized if the vector
$\d$ is orthogonal to both $\L$ and $\l$. It is easy to verify
that the opposite is also valid: if these vectors are orthogonal,
then we necessarily have (\ref{special}).

This simple example gives nevertheless a three-dimensional (not
complanar) configuration, if the vector $\d$ is nonzero and the
vectors $\L$ and $\l$ are not parallel. In this case the points
$\ri$, $\rf$, $\rA$, $\rB$ and $\rzer$ do not lie in the same
plane.

If $\d=0$, we have a complanar motion, in fact a symmetrical case
of the configuration considered in Sect.~\ref{method}, when the
collision leads to exchange of the velocities between the
particle and its incarnation. If $\d\ne 0$, but the vectors $\L$
and $\l$ are parallel, a symmetrical case of another type of the
complanar motion is realized, so-called ``mirror exchange'' of
the velocities \cite{Ech91}. The components of the velocities
along the direction of $\L$ and $\l$ are exchanged in this case
while the components along the direction $\d$ are exchanged with
simultaneous change of the signs.

Let us discuss now the general solution of the problem, i.e.
attempt to find the parameters of the motion for arbitrary
vectors $\L$, $\l$ and $\d$.

According to what has been proposed above, we should find
$\cone$, $\ctwo$ and $\c$ from the equations (\ref{main-eq}) and
require that they satisfy the orthogonality conditions
(\ref{orthog}). This will give the equations for the unknowns
$T_b$, $T_w$. Solution of these equations will determine the
configuration of the motion completely.

For convenience, we shall introduce the dimensionless unknowns
instead of $T_b$, $T_w$:
$$
\lb=-\frac{T_b}{T}, \quad \lw=\frac{T_w}{\tau}.
$$
Then Eq.~(\ref{main-eq}) gives for the velocities expressed
through these scalar unknowns
\ba\label{solut-c}
\theta\,\c &=& \lb\L+\lw\l+\d,\nonumber\\
\theta\,\cone &=& \left( \lone-\frac{\tau}{T}\lw^2 \right) \L
  + \lb\lw\l + \lb\d, \nonumber\\
\theta\,\ctwo &=& \lb\lw\L
  + \left( \ltwo-\frac{T}{\tau}\lb^2 \right) \L + \lw\d
\ea
where it is denoted
$$
\lone=1 + \frac{\tau}{T}, \quad \ltwo = 1 + \frac{T}{\tau},
\quad
\theta = T + \tau - T\lb^2 - \tau\lw^2.
$$

The orthogonality condition (\ref{orthog}) imposed on the vectors
(\ref{solut-c}) gives, after some algebra,
\be\label{lambda-eq}
(\d+\lb\L+\lw\l)\left(\d+\frac{\lone}{\lb}\L\right)=0, \quad
(\d+\lb\L+\lw\l)\left(\d+\frac{\ltwo}{\lw}\L\right)=0.
\ee

This algebraic equations for $\lb$, $\lw$ could be in principle
solved (they may be reduced to a third-degree equation for one
unknown). With the help of Eq.~(\ref{rzer-eq}), this would result
in the complete solution of the problem. Not trying to obtain
this solution in an explicit form, we shall discuss the question
of its existence.

It may be shown that, for arbitrary vectors $\L$, $\l$ and $\d$,
a unique pair of real numbers $\lb$, $\lw$ exists satisfying the
equations (\ref{lambda-eq}). However not all of these formal
solutions give a solution of the physical problem. The reason is
that, as a consequence of a natural inequalities
$$
t_i < t_0 < t_f, \quad t_A < t_0 < t_B,
$$
the following restrictions exist for the unknowns:
\be\label{lambda-restr}
|\lb| < 1, \quad |\lw| < 1.
\ee

The formal solution of Eq.~(\ref{lambda-eq}) satisfies this
restriction if the projection $\d_{{\mathrm proj}}$ of the vector
$\d$ onto the plane of the vectors $\L$, $\l$ is small enough,
but in general case this may be not valid.

\begin{figure}
%%Begin InstantTeX Picture
\let\picnaturalsize=N
\def\picsize{9cm}
\def\picfilename{les3.eps}
%If you do not have the picture file add:
%\let\nopictures=Y
%to the beginning of the file.
\ifx\nopictures Y\else{\ifx\epsfloaded Y\else\input epsf \fi
\let\epsfloaded=Y
\centerline{\ifx\picnaturalsize N\epsfxsize \picsize\fi \epsfbox{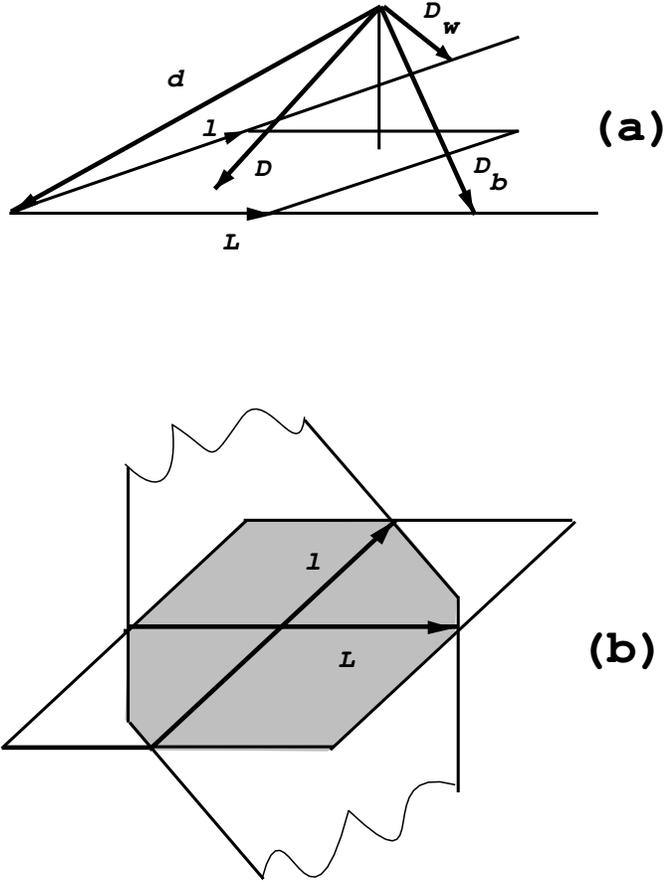}}}\fi
%%End InstantTeX Picture
\caption{\rm A self-collision with one passage through the
wormhole. (a)~The unknown parameters $\lb$, $\lw$ are determined
by the condition that $\D$ is orthogonal to $\D_b$ and $\D_w$.
(b)~The conditions $|\lb|<1$, $|\lw|<1$ (providing a correct
physical interpretation) are satisfied if the vector $\d_{{\mathrm
proj}}$ belongs to the intersection $\Omega$ of the two
parallelograms constructed with the help of the vectors $\L$ and
$\l$.}
\label{sol-exist}
\end{figure}

Indeed, Eq.~(\ref{lambda-eq}) means that among the three vectors
$$
\D=\d+\lb\L+\lw\l, \quad
\D_b=\d+\frac{\lone}{\lb}\L, \quad
\D_w=\d+\frac{\ltwo}{\lw}\L
$$
the first one should be orthogonal to the two others. These
vectors are drawn in Fig.~\ref{sol-exist}a. Geometrical analysis
of this figure shows that the inequalities (\ref{lambda-restr})
are fulfilled (and therefore the physical solution of the problem
exists) if the vector $\d_{{\mathrm proj}}$ belongs to some region
$\Omega$ near zero.

The region $\Omega$ is shown in Fig.~\ref{sol-exist}b where a
part of the plane $\L$, $\l$ is drawn. This region coincides with
the intersection of two parallelograms determined by the vectors
$\L$, $\l$ and having mutually perpendicular sides.

The condition $\d_{{\mathrm proj}}\in\Omega$ is sufficient for
existing a configuration with a self-collision and one passage
through the wormhole. It is difficult to formulate a sufficient
and necessary condition for this.

\section{Multiple  wormhole traversals}\label{multiple-pass}

We considered in much detail the scheme of motion with a
self-collision and one passage through the time machine (wormhole
traversal). However this is not the only possibility. Multiple
passages with a self-collision were discussed in \cite{Ech91}.
Only complanar or very close to complanar motions were considered
in this paper. This means that the space positions ($\ri$, $\rf$,
$\rA$, $\rB$, $\rzer$) of all key events lied in the same plane.
We shall consider the general situation (but with a point-like
body instead of a finite billiard ball).

\begin{figure}
%%Begin InstantTeX Picture
\let\picnaturalsize=N
\def\picsize{7cm}
\def\picfilename{les4.eps}
%If you do not have the picture file add:
%\let\nopictures=Y
%to the beginning of the file.
\ifx\nopictures Y\else{\ifx\epsfloaded Y\else\input epsf \fi
\let\epsfloaded=Y
\centerline{\ifx\picnaturalsize N\epsfxsize \picsize\fi \epsfbox{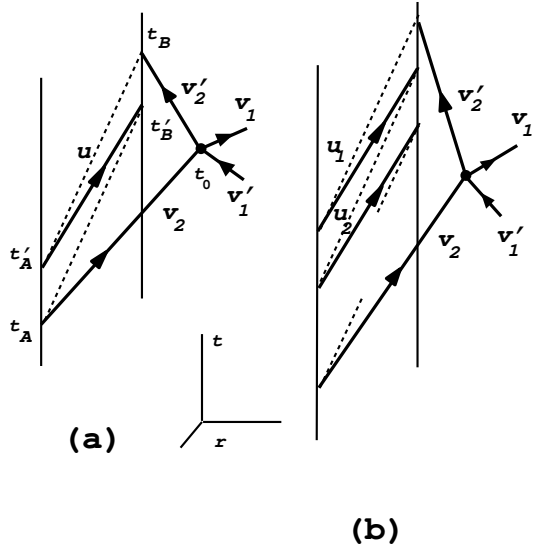}}}\fi
%%End InstantTeX Picture
\caption{\rm Double (a) and multiple (b) wormhole traversals
with a self-collision.Velocity $|\u|$ is less than $|\d|/\tau$,
thus repeated passages through the wormhole bring the
particle into the past.}
\label{fig-mult-pas}
\end{figure}

The scheme of consideration is quite analogous to that proposed
in Sects.~\ref{method},~\ref{one-travers}. The kinematical scheme
for a double passage is drawn in Fig.~\ref{fig-mult-pas}a. The
action has in this case the form differing from
Eq.~(\ref{action-v}) only by an additional term corresponding to
an additional passage through the time machine:
\be\label{action-2pass}
S= \frac{m}{2}\left[ \vone^2(t_0-t_i) + \vone'^2(t_f-t_0)
+\vtwo^2(t_0-t_A) +\vtwo'^2(t'_B-t_0)+\u^2(t'_B-t'_A)\right].
\ee
The following notations are used here:
\be\label{vel-events-2p}
\vone=\frac{\rzer-\ri}{t_0-t_i}, \quad
\vone'=\frac{\rf-\rzer}{t_f-t_0}, \quad
\vtwo=\frac{\rzer-\rA}{t_0-t_A}, \quad
\vtwo'=\frac{\rB-\rzer}{t_B-t_0}, \quad
\u=\frac{\r'_B-\r'_A}{t'_B-t'_A}.
\ee

Variation of this action in the parameters $\rzer$, $t_0$,
$t_A+t'_B$, $t'_A+t_B$ gives the equations
\be\label{vel-eq-2p}
\vone + \vtwo = \vone' + \vtwo', \quad
\vone^2+\vtwo^2 = \vone'^2+\vtwo'^2, \quad
\vtwo'^2 = \u^2=\vtwo^2
\ee
(the momentum and energy conservation and the rule `the
in-velocity is equal to the out-velocity'). The equalities
(\ref{vel-eq-2p}) may be considered as equations for the unknown
parameters of the motion: $\rzer$, $t_0$, $t_A+t'_B$, $t'_A+t_B$.
The concrete scheme of their solution, quite similar to one
proposed in Sect.~\ref{one-travers}, may be developed. It reduces
the problem to the solution of a system of algebraic equations.

Now we can easily go over to the case of an arbitrary number $N$
of passages through the time machine. The scheme of such a motion
is presented in Fig.~\ref{fig-mult-pas}b. It is evident how the
action for this motion may be written and what is the result of
its variation. In the notations evident from the figure, we have
\be\label{vel-eq-Np}
\vone + \vtwo = \vone' + \vtwo', \quad
\vone^2+\vtwo^2 = \vone'^2+\vtwo'^2, \quad
\vtwo'^2 = \u_1^2=\u_2^2=\dots=\u_N^2=\vtwo^2.
\ee

\begin{figure}
%%Begin InstantTeX Picture
\let\picnaturalsize=N
\def\picsize{7 cm}
\def\picfilename{les5.eps}
%If you do not have the picture file add:
%\let\nopictures=Y
%to the beginning of the file.
\ifx\nopictures Y\else{\ifx\epsfloaded Y\else\input epsf \fi
\let\epsfloaded=Y
\centerline{\ifx\picnaturalsize N\epsfxsize \picsize\fi \epsfbox{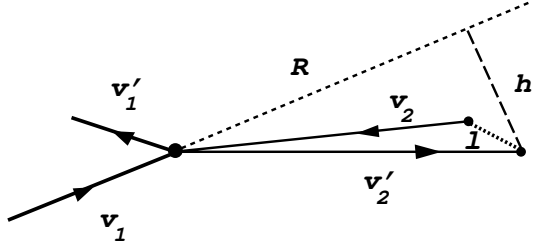}}}\fi
%%End InstantTeX Picture
\caption{\rm Many wormhole traversals with the collision far from
the mouths.}
\label{example-2p} \end{figure}

To demonstrate existence of a solution to these equations, we
shall consider a concrete example, suggesting that the collision
occurs far from the wormhole (see Fig.~\ref{example-2p}). Suppose
that the following inequalities are valid:
$$
R\gg h \gg l
$$
where $R$ is the distance, along the initial trajectory, from the
collision point to the point closest to the mouth $B$, $h$ is the
distance of the mouth $B$ from the initial trajectory, and
$l=|\l|$ the distance between the mouths.

As a consequence of the condition $R\gg l$, the velocity $v'_2$
with which the ball (particle) heads towards the wormhole and the
velocity $v_2$ with which it returns are very nearly
antiparallel. Using this fact and the rule $v_2=v'_2$, it is easy
to demonstrate that
$$
v_2=v'_2=v_1 \cos{\frac{h}{R}} =
v_1 \left[ 1- \frac12\left(\frac{h}{R} \right)^2\right],
$$
where $v_1$ is the initial velocity of the particle.

After this, the simple argument based on the kinematical scheme
Fig.~\ref{fig-mult-pas}b leads to the following relation for the
number $n$ of the wormhole traversals:
$$
n=\frac{2R/l}{v_1\frac{\tau}{l}
\left[1- \frac12 \left( \frac{h}{R}\right)^2\right]-1}.
$$

Thus, if $v_1> l/\tau$, there are infinite number of possible
configurations of the motion for each initial trajectory. When
$R$ tends to infinity, the number of traversals $n$ also tends to
infinity. The ball may pass through the time machine many times
provided the collision with the incarnation occurs far from the
wormhole. There is a minimal distance
$$
R_{\mathrm min}=\frac{h}{\sqrt{2\,\left( 1- \frac{l}{v_1\tau}\right)}}
$$
for which there are permitted trajectories. While $R$ decreases
from infinity to this $R_{\mathrm min}$, the number $n$ decreases
from infinity to some minimal value and after this increases to
infinity again.

The inequality $v_1 < l/\tau$ leads to negative $n$ for any $R>h$
that is impossible. This means that self-collision in the
considered configuration is impossible for such a small initial
velocity of the particle.

Besides the configurations already discussed, one could in
principle consider configurations with multiple self-collisions.
Multiple wormhole traversals are also necessary to provide
multiple self-collisions of the same particle. The equations for
such configurations may be written down and analyzed in
essentially the same way as it has been done in the case of one
collision. However the analysis is in this case much more
complicated. A preliminary analysis carried out for the case of
two self-collisions (and two transversals) allows one to suggest that 
two-collision configurations, under the conditions $\rho\ll r\ll l$ 
(see Sect.~\ref{intro}) are impossible. However the question requires 
further investigation.

\section{Jinnee of Time Machine}\label{jinn}

We have systematically considered different configurations of
motion of a particle (point-like body) with traversals the time
machine and a self-collision. It is a high time now to remark
that each of these configurations may be interpreted in different
ways.

Begin with the simplest configuration Fig.~\ref{traject}b. We
supposed so far that the particle moved along the line 1, then,
after collision, continued the motion along the line $2'$, then,
on escaping from the wormhole, went along the line 2, and after
the self-collision in the event $O$ finished the motion along the
line $1'$.

A different interpretation of the same dynamical scheme suggests
that the particle moves only along the lines 1 and $1'$. In the
event $O$ it collides not with itself, but with another particle
having the same mass and interaction (identical to the first
one). The second particle emerges from the mouth $A$, travels
along the line 2 to the event of collision $O$ and then, after
the collision, returns to the wormhole, passing along $2'$ and
entering the mouth $B$.

In this interpretation there are two particles. The particle 1
never enters the time machine, and the particle 2 goes out of the
time machine only to collide with the particle 1 and deflect it
from the direct line of motion. After this the second particle
returns to the wormhole. Such a particle has a closed time-like
trajectory (world line). It is nothing else than Jinnee of Time
Machine discussed in \cite{jinn1,jinn2} (see \cite{MenNov} for
the discussion of a quantum version of Jinnee).

It is important that all characteristics of the particle 2 should
be identical to those of the particle 1. All parameters of the
motion are then quite the same for both interpretations of it. Of
course, if finite dimension of the ball is taken into account,
then the motions corresponding to the two interpretations turn
out to be slightly different. For example, positions of the two
colliding balls are different in the cases of the first and
second interpretations (without or with the Jinnee).

It is interesting to consider the Jinnee interpretation for a
special case of the same configuration, namely for the motion
presented in Fig.~\ref{trivial}. In the usual interpretation
(accepted in the preceding sections) the particle, because of
the collision, is deflected from the direct motion and enters the
wormhole. Escaping from the wormhole, it collides itself,
providing deflection of its first incarnation, and proceeds
further to the final event $f$. The motion is rather
complicated.

The Jinnee interpretation of the same motion is quite trivial.
The particle freely moves along the direct line from the event
$i$ to the event $f$ without any collision. Simultaneously the
second particle (Jinnee) exists. It goes out of the wormhole only
to move along the direct line from $A$ to $B$ and return to the
wormhole. It does not collide the first particle, but it passes
very nearly to it (that is possible in the approximation of
point-like bodies and short-range interaction between them).

Thus, in the configuration Fig.~\ref{trivial} we have
alternatively non-trivial motion of a single particle or trivial
motion of two particles, one of which is Jinnee. In the general
case Fig.~\ref{traject}b both interpretations (without Jinnee or
with Jinnee) lead to non-trivial motions.

The preceding argument evidently applies to a similar, but more
complicated configuration presented in Fig.\
\ref{fig-mult-pas}ab. Again we may interpret the motion i)~either
as a motion of a single particle entering the wormhole and
undergoing self-collision, ii)~or as a motion of the particle
travelling outside the wormhole, but colliding with Jinnee of
Time Machine. Now we have Jinnee that passes through the wormhole
two or several times. Usually Jinnee travels freely outside the
wormhole. However once, being outside the wormhole, Jinnee goes
far from it, collides the by-passing particle and returns to the
wormhole.

Thus, when possible existence of bodies with closed time-like
world lines (trajectories) is taken into account, different
interpretations may be given for each concrete configuration of
the motion with a self-collision and traversals of the time
machine. From a certain point of view, interpretations with
Jinnee are simpler than those with no Jinnee.\footnote{By the
way, this is reflected in the fact that the notations for
velocities chosen in Sect.~\ref{one-travers} only for
mathematical convenience, turned out to correspond to the Jinnee
interpretation rather than to the no-Jinnee one.}

Since different interpretations are dynamically quite
equivalent\footnote{in the approximation of point-like bodies},
one of them may be taken for the analysis arbitrarily. For
example, when we need to specify all configurations characterized
by a definite number of traversals, it is convenient to look for
those configurations where the particle never enters the wormhole
but collides with Jinnee. Afterwards each configuration found in
such a way may be interpreted in terms of a single particle, with
no Jinnee.

We should underline that different interpretations are
dynamically equivalent only in the approximation of point-like
bodies. For small balls instead of points, slight differences
arise in mutual positions these balls have in different
interpretations.

Moreover, the interpretations with Jinnee seem to be
much more natural for simple objects such as elementary
particles, than for complicated bodies such as billiard balls.
One may suppose that Jinnee-interpretations may become physically
significant only for elementary particles. In this case existence
of completely identical objects is not at all astonishing.
The objects identical to one under investigation but moving along
closed world lines, evidently resemble vacuum loops arising in
quantum theory.

Thus, classical theory of bodies moving in presence of a time
machine naturally leads to some concepts, otherwise arising only
in quantum theory. This property seems very interesting and
deserving further exploration.

\section{Conclusion}\label{conclus}

We considered the motion (including a self-collision) of a
point-like body in the time-machine spacetime. The set of
configurations of such a motion turned out to be richer than it
has been known earlier. The approximation of a point-like body
and the method of minimal action proved to be efficient for
investigating three-dimensional configurations of the motion.

We found main features of such motions (for a specific model of a
time machine) and found that, given boundary conditions (initial
%%%% and final positions ad times), %%%%
and final positions and times), 
many configurations of motion are
possible, differing by the number of traversals through the time
machine. An interesting feature of these configurations is
discovered: they may be interpreted in different ways if the
bodies with closed time-like world lines (Jinnee of Time Machine)
are introduced. There is a certain analogy between (classical)
trajectories of Jinnee and vacuum loops typical for relativistic
quantum theory.

With all this taken into account, it seems interesting to
consider scattering of a particle on the time machine and the
role played in this process by different configurations of
motion. It may be supposed that correct formulation of this
problem may be given only in the framework of quantum theory.

\vspace{0.5cm}
\centerline{\bf ACKNOWLEDGMENTS}

One of the authors (M.M.) is indebted to J.~Audretsch and
H.~v.~Borzeszkowski for useful discussions. The work was
supported in part by the Danish National Research Foundation
through the establishment of the Theoretical Astrophysics Center,
by the Danish Natural Science Research Council, grant 11-9640-1,
the Russian Foundation for Fundamental Research, grant
95-01-00111a, and the Deutsche Forschungsgemeinschaft.

\end{document}